\documentclass{article}%
\usepackage{amssymb}
\usepackage{epsf}%
\usepackage{amsmath}%
\usepackage{amsfonts}%
\usepackage{graphicx}
\begin{document}

\title{Principal null directions of perturbed black holes}
\author{Z. Perj\'{e}s and M. Vas\'{u}th\\KFKI Research Institute for Particle and Nuclear Physics,\\H--1525, Budapest 114, P.O.B.\ 49, Hungary}
\maketitle

\begin{abstract}
The properties of principal null directions of a perturbed black hole are
investigated. It shown that principal null directions are directly observable
quantities characterizing the space-time. A definition of a perturbed
space-time, generalizing that given by Stewart and Walker is proposed. This
more general framework allows one to include descriptions of a given
space-time other than by a pair $(M,g)$ where $M$ is a four-dimensional
differential manifold and $g$ a Lorentz metric. Examples of alternative
characterizations are the curvature representation of Karlhede and others, the
Newman-Penrose representation or observable quantities involving principal
null directions. The conditions are studied under which the various
alternative choices of observables provide equivalent descriptions of the space-time.

\end{abstract}

\section{Introduction}

The necessary and sufficient condition for a space-time to be algebraically
special in the Petrov classification is \cite{Kramer} that the curvature
invariants $I$ and $J$ satisfy
\begin{equation}
I^{3}-27J^{2}=0.\label{algsp}%
\end{equation}
The curvature invariants are the determinantal expressions in the components
of the Weyl spinor \cite{NP}
\begin{align}
I  & =\left\vert
\begin{array}
[c]{cc}%
\Psi_{0} & \Psi_{1}\\
\Psi_{3} & \Psi_{4}%
\end{array}
\right\vert +3\left\vert
\begin{array}
[c]{cc}%
\Psi_{2} & \Psi_{1}\\
\Psi_{3} & \Psi_{2}%
\end{array}
\right\vert \label{I}\\
J  & =\left\vert
\begin{array}
[c]{ccc}%
\Psi_{0} & \Psi_{1} & \Psi_{2}\\
\Psi_{1} & \Psi_{2} & \Psi_{3}\\
\Psi_{2} & \Psi_{3} & \Psi_{4}%
\end{array}
\right\vert .\nonumber
\end{align}
A Newman-Penrose (NP) notation for the components of a symmetric spinor is
used here, \textit{i.e.},
\begin{equation}
\Psi_{m}=\Psi_{0...1}\label{notpsi}%
\end{equation}
with $m$ the number of spinor indices $1.$

Baker and Campanelli \cite{Baker} introduce the \textit{speciality index} of
the space-time as the ratio
\begin{equation}
S=\frac{27J^{2}}{I^{3}}.\label{spec}%
\end{equation}
For a Kerr-Newman black hole in the Kinnersley tetrad \cite{Kinnersley}, the
only nonvanishing component of the Weyl spinor is $\Psi_{2}.$ Expressions
(\ref{I}) then take simple forms and the condition (\ref{algsp}) is satisfied
(as it should for a type D space-time). Baker and Campanelli go on and expand
the speciality index in powers of an arbitrary perturbation parameter
$\varepsilon$ of the black hole:
\begin{equation}
S=1-3\varepsilon^{2}\frac{\Psi_{0}\Psi_{1}}{\left(  \Psi_{2}\right)  ^{2}%
}+\mathcal{O}\left(  \varepsilon^{3}\right)  .
\end{equation}
This is a perplexing result since it implies that perturbed black holes, in
the first-order approximation, are algebraically special. There exists an
extensive literature \cite{Couch,Wald,Chandrasekhar,Brink} of `algebraically
special perturbations'. What \textit{are} these then? It is our purpose in the
present paper to shed some light on this apparent controversy by examining the
properties of the principal null directions. The clarification of this point
leads one to fundamental issues such as the notion of the perturbation of a space-time.

A general framework for perturbations of space-times has been discussed by
Stewart and Walker \cite{Stewart}. They propose the following

\textbf{Definition 1 (Stewart-Walker)} \textit{A space-time $\left(
\mathcal{M}^{\prime},g^{\prime}\right)  $ consisting of the manifold
$\mathcal{M}^{\prime}$ and metric $g^{\prime}$ is a perturbation of some given
space-time $\left(  \mathcal{M},g\right)  $ if there exists a smooth
one-parameter family of space-times $\left(  \mathcal{M},g\left(
\lambda\right)  \right)  $ in a Hausdorff five-dimensional manifold
$\mathbf{M}^{(5)}$ which contains $\left(  \mathcal{M},g\right)  =\left(
\mathcal{M},g\left(  0\right)  \right)  $ and $\left(  \mathcal{M}^{\prime
},g^{\prime}\right)  =\left(  \mathcal{M},g\left(  \varepsilon\right)
\right)  .$ The unperturbed or background space-time $\left(  \mathcal{M}%
,g\right)  $ is given by $\lambda=0.$ In the parameter interval $\lambda
\in\left(  -\varepsilon,\varepsilon\right)  ,$ the $g^{ab}\left(
\lambda\right)  $ define a piecewise smooth tensor field $g^{\alpha\beta}$ on
$\mathbf{M}^{(5)}$ of signature $\left(  0+---\right)  .$ The singular
hypersurfaces $g^{\alpha\beta}\left(  d\lambda\right)  _{\beta}=0$ are then
the original space-time manifolds $\mathcal{M}_{\lambda}$. }

A weakness of this definition is that it is too restrictive to be applicable
to some important works in perturbation theory. Many treatments of black-hole
perturbations characterize the space-time by using curvature components and
other quantities which are not considered in the definition. As an example,
Chandrasekhar \cite{Chandra2} characterizes a perturbed black hole such that
the dyad components of the Weyl spinor plus the optical scalars differ only by
first-order terms from their form in the background space-time. Such
treatments are admissible only in a general framework in which the
characterization of the space-time in more than one ways is allowed. For
example, in the curvature representation \cite{Karlhede}, a space-time is
locally completely determined by the Riemann tensor and a finite number of its
derivatives in a moving frame.

It thus appears necessary to formulate the perturbation problem in a more
general way than by the Stewart-Walker definition. If one wants to encompass
the various alternative characterizations of the space-time, one may replace
the pair $\left(  \mathcal{M}^{\prime},g^{\prime}\right)  $ in the definition
by a pair $\left(  \mathcal{M}^{\prime},\mathbf{O}^{\prime}\right)  $ where
$\mathbf{O}^{\prime}$ is some complete set of measurable quantities
characterizing the space-time. The complete characterization of the geometry
is the central issue in what is known  the equivalence problem
\cite{Karlhede,Michael}.  

\textbf{Definition 2} \textit{ A set of observables }$\mathbf{O}$ \textit{on
some open set }$\mathcal{U}$\textit{ is said to locally completely
characterize the space-time containing }$\mathcal{U}$ \textit{if the
differentiable and metric\ structures are fixed uniquely on }$\mathcal{U}%
$\textit{ by the values of }$\mathbf{O.}$

Obvious examples of the choice of $\mathbf{O}^{\prime}$ are the perturbed
metric $g^{\prime}$ or the quantities used in the curvature representation
\cite{Karlhede}. An incomplete set of observables, for example, in a
space-time containing a number of different matter fields is obtained when
dropping the subset of observables such as the field stresses of one of the
fields from among the observables. 

The proposed new definition of a perturbed space-time raises some questions
which did not complicate the old definition. An important such question is
whether two different choices of $\mathbf{O}^{\prime}$ yield equivalent
treatments of the perturbation problem. In the context of linear
perturbations, we may formulate this as the requirement that the two choices
$\mathbf{O}$ and $\mathbf{O}^{\prime}$ of the observables are related by
$\mathbf{O}^{\prime}=M\mathbf{O}$ where $M$ is a nonsingular matrix.

Our goal in this paper is to investigate the sets of observable quantities
characterizing the space-time in order to clarify if different choices of
these sets of observables will yield equivalent descriptions of the
perturbations. In particular, we want to clarify how the various choices of
observable quantities behave under perturbations of the system. As a first
step on this route, in the next section we recapitulate the behavior of
rotational metric perturbations in Hartle's theory \cite{Hartle}. In Sec.
\ref{pndir}, we briefly recall the concept of principal null directions of a
symmetric spinor. The application of this general theory to the
electromagnetic and gravitational fields in a Kerr-Newman black hole is
reviewed in Sec. \ref{knbh}. The behavior of principal null directions of the
electromagnetic vs. gravitational field to first order in the perturbation
parameter is worked out in Secs. \ref{empert} and \ref{gravpert}, respectively.

We find, as an unexpected result of this investigation, that there are certain
choices of the sets of observables which provide inequivalent descriptions of
the space-time. The existence of these inequivalent sets is shown to have
important consequences for the theory of gravitational radiation. This
finding, when combined with the properties of principal null directions in the
perturbative picture, makes it possible to reach the correct interpretation of
`algebraically special perturbations'.

\section{Illustrative example: rotational perturbations\label{rotpert}}

In the Hartle theory \cite{Hartle} of rotational perturbations, the metric of
the unperturbed vacuum is given by the Schwarzschild solution with mass $m.$
The perturbed space-time has the metric
\begin{align}
ds^{2}  & =\left(  1+2h\right)  \left(  1-\frac{2m}r\right)  dt^{2}-\left(
1+2j\right)  \left(  1-\frac{2m}r\right)  ^{-1}dr^{2}\nonumber\\
& -\left(  1+2k\right)  r^{2}\left[  d\vartheta^{2}+\sin^{2}\vartheta\left(
d\varphi-\omega dt\right)  ^{2}\right]  .
\end{align}
Here the perturbation functions $h,$ $j,$ $k$ and $\omega$ are assumed to
depend smoothly on the perturbation parameter, \textit{i.e.}, the angular
velocity $\Omega$. Additional information on the form of these quantities is
obtained from the discrete symmetries of the rotating system. Taking into
account that the simultaneous reversal of time direction and the reversal of
the sense of rotation is an exact symmetry, we have that the series expansion
of the function $\omega$ contains only odd powers of the angular velocity. The
expansion of all other perturbation functions has only even powers in $\Omega
$. Hence one finds that the only unknown function in the perturbation problem
to first order in $\Omega$ is $\omega$. In fact, the field equations yield an
uncoupled, second-order linear differential equation for the function
$\omega.$ Given the solution of this equation, one may proceed to solve the
perturbation problem to second order in the angular velocity $\Omega$. In the
second approximation, the field equations are coupled second-order linear
differential equations for the remaining perturbation functions. The function
$\omega$ contributes quadratic source terms in these equations.

The rotational perturbations of the Schwarzschild black hole illustrate our
point: a description which takes into account the full physical properties of
the perturbed system can yield restrictions on the form of the series
expansion in the perturbation parameter.

\section{Principal null directions\label{pndir}}

The principal spinors of a $k$-index symmetric spinor $\varphi_{ABC...K}$ are
defined as the nonzero spinors $o^{A}$ satisfying the condition
\begin{equation}
\varphi_{ABC...K}o^{A}o^{B}o^{C}...o^{K}=0.\label{eq1}%
\end{equation}
Let $o^{0}$ be the nonzero component and let us introduce the complex ratio
\begin{equation}
z=\frac{o^{1}}{o^{0}}\ .\label{eq2}%
\end{equation}
The roots of the complex algebraic equation
\begin{equation}
\varphi_{0}+k\varphi_{1}z+\left(
\begin{array}
[c]{c}%
k\\
2
\end{array}
\right)  \varphi_{2}z^{2}+\left(
\begin{array}
[c]{c}%
k\\
3
\end{array}
\right)  \varphi_{3}z^{3}+...+\varphi_{k}z^{k}=0
\end{equation}
[where we are using the notation (\ref{notpsi}) for the components of the
spinor] define the flagpoles \cite{PR} $o^{A}\overline{o}^{A^{\prime}}.$ The
principal null directions of $\varphi_{ABC...K}$ are represented, up to real
multiplying factors, by these flagpoles. By the fundamental theorem of the
algebra, there exist $k$ roots.

The various coincidences of the principal null directions are classified by
the partitions of the number $k$. In the generic type, there are $k$ distinct
principal directions, represented by $k$ vectors of the light cone in the
tangent space $T_{p}$ at point $p$ of the space-time (Figure 1). All other
types are called algebraically special.
%TCIMACRO{\FRAME{ftbpFU}{5.0211in}{2.4025in}{0pt}{\Qcb{{}The angle subtended by
%two principal null directions for an observer is given by the line element ds
%of his celestial sphere.}}{}{pnd2.eps}{\special{ language "Scientific Word";
%type "GRAPHIC";  maintain-aspect-ratio TRUE;  display "USEDEF";
%valid_file "F";  width 5.0211in;  height 2.4025in;  depth 0pt;
%original-width 21.8824in;  original-height 10.4089in;  cropleft "0";
%croptop "1";  cropright "0.9995";  cropbottom "0";
%filename 'E:/tmp/Pnd2.eps';file-properties "XNPEU";}}}%
%BeginExpansion
\begin{figure}
[ptb]
\begin{center}
\includegraphics[
trim=0.000000in 0.000000in 0.010942in 0.000000in,
height=2.4025in,
width=5.0211in
]%
{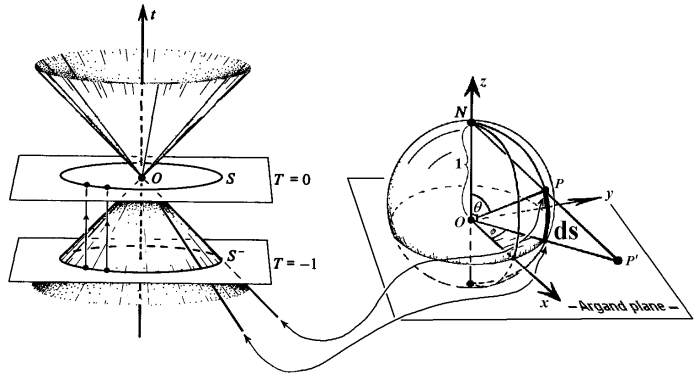}%
\caption{{}The angle subtended by two principal null directions for an
observer is given by the line element ds of his celestial sphere.}%
\end{center}
\end{figure}
%EndExpansion

Principal null directions at any given point are \emph{directly observable
quantities} of the space-time. An observer with world line through the point
$p$ can detect principal null directions by looking at a distant sphere
\cite{Gunnarsen}. In the lowest-order approximation, the observed image of the
sphere is distorted to an ellipse. Kristian and Sachs \cite{Kristian}
characterize the distortion by the ratio of the major and minor axes of the
ellipse:
\begin{equation}
D=\frac{1}{2}C\left(  R,\vartheta,\varphi\right)  R^{2}%
\end{equation}
depending both on the distance $R$ and the direction $\left(  \vartheta
,\varphi\right)  $ of the source. To the present accuracy, the distance $R$
can be equivalently taken to be the luminosity distance, the distance by the
apparent size or area distance. The coefficient $C\left(  R,\vartheta
,\varphi\right)  $ is the projection of the Weyl tensor in the tangent plane
of the celestial sphere \cite{Kristian}. At the principal null directions, the
coefficient $C$ of the distortion has the limiting value $\lim_{R=0}C\left(
R,\vartheta,\varphi\right)  =0.$

The angle subtended by a pair of these directions is given as the invariant
length $s$ on the unit sphere $\mathcal{S}^{2}$ with metric
\begin{equation}
ds^{2}=d\vartheta^{2}+\sin^{2}\vartheta d\varphi^{2}.
\end{equation}
In an orthogonal frame $\{\mathbf{t,x,y,z }\}$, the position of point $p$ of
Minkowski space-time is $T\mathbf{t}+X\mathbf{x}+Y\mathbf{y}+Z\mathbf{z}$. The
sphere $\mathcal{S}^{2}$ is represented \cite{PR} by the intersection of the
past null cone with the hyperplane $T=-1$. The polar coordinates $\vartheta$
and $\varphi$ are related to the complex stereographic coordinates $z$ and
$\overline{z}$ by the transformation
\begin{equation}
z=e^{i\varphi}\cot\frac\vartheta2.
\end{equation}
The metric of $\mathcal{S}^{2}$ in stereographic coordinates has the form
\begin{equation}
ds^{2}=\frac4{\left(  1+z\overline{z}\right)  ^{2}}dzd\overline{z}.
\end{equation}
The coordinate system $\left(  z,\overline{z}\right)  $ is singular at the
north pole where $z\rightarrow\infty,$ and the coordinate distortion due to
the presence of the metric coefficient $P=\frac1{2^{1/2}}\left(
1+z\overline{z}\right)  $ increases indefinitely as one approaches the pole.

There exists a sizable literature of the properties of principal null
directions. Penrose and Rindler devote a chapter of their monograph \cite{PR}
to a detailed study of this subject. Gunnarsen et al. \cite{Gunnarsen} develop
a numerical approach, based on the d'Inverno-Russel Clark version of the
Ferrari formula, for computing the gravitational principal null directions.
They employ this method for the Kastor-Traschen metrics containing colliding
black holes. The status of observations of principal null directions, along
with various proposed techniques has been reviewed by Chrobok and Perlick
\cite{Chrobok}.

\section{The Kerr-Newman black hole\label{knbh}}

The Kerr-Newman black hole with mass $\mathrm{m}$, rotation parameter $a$ and
electric charge $\mathrm{e}$ has the metric
\begin{align}
ds^{2}  & =\left(  1-\frac{2\mathrm{m}r-\mathrm{e}^{2}}{\zeta\overline{\zeta}%
}\right)  \left(  dt-a\sin^{2}\vartheta d\varphi\right)  ^{2}%
\nonumber\\
& +2\left(  dt-a\sin^{2}\vartheta d\varphi\right)  \left(  dr+a\sin
^{2}\vartheta d\varphi\right) \label{ds2}\\
& -\zeta\overline{\zeta}\left(  d\vartheta^{2}+\sin^{2}\vartheta d\varphi
^{2}\right) \nonumber
\end{align}
with the four-potential
\begin{equation}
A=-\frac{\mathrm{e}r}{\zeta\overline{\zeta}}\left(  dt-a\sin^{2}\vartheta
d\varphi\right) \label{A}%
\end{equation}
where
\begin{equation}
\zeta=r-ia\cos\vartheta.
\end{equation}
The coordinates $t$ and $r$ run from $-\infty$ to $\infty$ while $\vartheta$
and $\varphi$ are coordinates on a 2-sphere such that $\varphi$ is periodic
with period $2\pi$ and $\vartheta$ ranges from $0$ to $\pi$. In the generic
case, the expression
\begin{equation}
{\mathbf{\Delta}}=r^{2}+a^{2}-2\mathrm{m}r+\mathrm{e}^{2} \
\end{equation}
has two distinct roots.

The two double gravitational principal null directions in the Kinnersley
tetrad are at the south versus the north poles of $\mathcal{S}^{2}$,
\textit{i.e.}, at the values of the stereographic coordinates $z=0$ and
$z=\infty.$ Since the coordinate system is singular at the north pole, it is
more advantageous to use the null tetrad \cite{NP}
\begin{align}
D  & \equiv\ell^{a}\frac\partial{\partial x^{a}}=\frac\partial{\partial
r}\nonumber\\
\Delta & \equiv n^{a}\frac\partial{\partial x^{a}}=\frac12\left(  \frac{2
\mathrm{m}r-\mathrm{e}^{2}}{\zeta\overline{\zeta}}-1\right)  \frac
\partial{\partial r}+\frac\partial{\partial t}\label{tetrad}\\
\delta & \equiv m^{a}\frac\partial{\partial x^{a}}=\frac1{2^{1/2}%
\overline{\zeta}}\left[  ia\sin\vartheta\left(  \frac\partial{\partial
t}-\frac\partial{\partial r}\right)  +\frac\partial{\partial\vartheta}+\frac
i{\sin\vartheta}\frac\partial{\partial\varphi}\right] \nonumber\\
\overline{\delta}  & \equiv\overline{m}^{a}\frac\partial{\partial x^{a}%
}.\nonumber
\end{align}
In the NP notation, the Maxwell tensor components are
\begin{align}
\Phi_{0}  & \equiv F_{ab}\ell^{a}m^{b}=0\nonumber\\
\Phi_{1}  & \equiv\frac12F_{ab}\left(  \ell^{a}n^{b}+\overline{m}^{a}%
m^{b}\right)  =\frac{\mathrm{e}}{2^{1/2}\zeta^{2}}\label{Phis}\\
\Phi_{2}  & \equiv F_{ab}\overline{m}^{a}n^{b}=\frac{i\mathrm{e}a\sin
\vartheta}{\zeta^{3}}.\nonumber
\end{align}
The significant components of the Weyl curvature $C_{abcd}$ are
\begin{align}
\Psi_{2}  & =\frac{\mathrm{e}^{2}-\mathrm{m}\overline{\zeta}}{\zeta
^{3}\overline{\zeta}}\nonumber\\
\Psi_{3}  & =-3ia\sin\vartheta\frac{\mathrm{m}\overline{\zeta}-\mathrm{e}^{2}%
}{2^{1/2}\zeta^{4}\overline{\zeta}}\label{Psis}\\
\Psi_{4}  & =3a^{2}\sin^{2}\vartheta\frac{\mathrm{m}\overline{\zeta
}-\mathrm{e}^{2}}{\zeta^{5}\overline{\zeta}}\nonumber
\end{align}
and the remaining two components vanish,
\[
\Psi_{0}\equiv-C_{abcd}\ell^{a}m^{b}\ell^{c}m^{d}=0\ , \quad\Psi_{1}%
\equiv-C_{abcd}\ell^{a}n^{b}\ell^{c}m^{d}=0 \ .
\]
The curvature singularities are at $r=0$ and $\vartheta=\pi/2$ .

Let the complex ratio of the Maxwell tensor components be denoted
\begin{equation}
w=\frac{\Phi_{2}}{2\Phi_{1}}=\frac{ia\sin\vartheta}{2^{1/2}\zeta}.
\end{equation}
In this notation, the Weyl tensor components satisfy
\begin{equation}
\Psi_{3}=3w\Psi_{2},\qquad\Psi_{4}=2w\Psi_{3}.
\end{equation}
The double gravitational principal null directions are at the finite
coordinate values $z=0$ and $z=-1/w.$

The observation of principal null directions in a Kerr-Newman black hole can
be accomplished in an alternative way \cite{MTW} to the one described in the
previous section. There exist special null congruences (`photon trajectories')
consisting of the integral curves of principal null directions, characterized
by vanishing first curvature $\kappa$ and shear $\sigma$. For these curves,
the polar angle $\vartheta$ is constant. The axis value of the angular
momentum is $L_{z}=aE\sin^{2}\vartheta,$ with $E$ the conserved energy of the
motion, and the Carter separation constant vanishes, $K=0$.

\section{Electromagnetic perturbations\label{empert}}

For later reference, we consider here the perturbations of the electromagnetic
field in the neighborhood of a black hole. Our first choice for the
(incomplete) set of observable quantities is the Maxwell field components. The
unperturbed field is given by the components (\ref{Phis}). The perturbed
Maxwell field has the expansion
\begin{align}
\Phi_{0}  & =\varepsilon\Phi_{0}^{(1)}+\mathcal{O}\left(  \varepsilon
^{2}\right) \nonumber\\
\Phi_{1}  & =\Phi_{1}^{\left(  0\right)  }+\varepsilon\Phi_{1}^{(1)}%
+\mathcal{O}\left(  \varepsilon^{2}\right) \\
\Phi_{2}  & =\Phi_{2}^{\left(  0\right)  }+\varepsilon\Phi_{2}^{(1)}%
+\mathcal{O}\left(  \varepsilon^{2}\right) \nonumber
\end{align}
where the parenthesized superscripts indicate the degree in the perturbation
parameter $\varepsilon$.

The principal null directions of the perturbed Maxwell field are given by the
solutions of the quadratic equation
\begin{equation}
z^{2}+a_{1}z+a_{0}=0
\end{equation}
where the coefficients are defined
\begin{equation}
a_{0}=\frac{\Phi_{0}}{\Phi_{2}},\qquad a_{1}=\frac{2\Phi_{1}}{\Phi_{2}}.
\end{equation}
A possible alternative choice of the observable quantities is the set of
principal null directions. We now want to examine if these two choices of the
sets of observables are equivalent. The relation between the coefficients
$a_{i}$ and the roots $z_{1}$ and $z_{2}$ is given by the Viet\'{e} formulae
\begin{align}
-a_{1}  & =z_{1}+z_{2}\label{Viete}\\
a_{0}  & =z_{1}z_{2}.\nonumber
\end{align}
Expanding all observable quantities in power series of $\varepsilon$ , we
have
\begin{align}
a_{0}  & =\varepsilon a_{0}^{(1)}+\mathcal{O}\left(  \varepsilon^{2}\right)
\nonumber\\
a_{1}  & =a_{1}^{(0)}+\varepsilon a_{1}^{(1)}+\mathcal{O}\left(
\varepsilon^{2}\right) \nonumber\\
z_{1}  & =\varepsilon z_{1}^{(1)}+\mathcal{O}\left(  \varepsilon^{2}\right) \\
z_{2}  & =z_{2}^{(0)}+\varepsilon z_{2}^{(1)}+\mathcal{O}\left(
\varepsilon^{2}\right) \nonumber
\end{align}
where the unperturbed principal null directions are at $z_{1}^{(0)}=0$ and
$z_{2}^{(0)}=-1/w$. Inserting in Eqs. (\ref{Viete}), the unperturbed terms
cancel and the first-order parts yield
\begin{align}
a_{1}^{(1)}  & =-\left(  z_{1}^{(1)}+z_{2}^{(1)}\right) \\
a_{0}^{(1)}  & =z_{1}^{(1)}z_{2}^{(0)}.\nonumber
\end{align}
From these relations we see that the set $\left(  z_{1}^{(1)},z_{2}%
^{(1)}\right)  $ of first-order observables is linearly equivalent to the set
$\left(  a_{0}^{(1)},a_{1}^{(1)}\right)  .$ Our conclusion is that the field
stresses of the perturbed Maxwell tensor yield a description equivalent with
the perturbed principal null directions.

\section{Gravitational perturbations\label{gravpert}}

In the literature of perturbed black holes \cite{Chandra2}, the sets of
observable quantities characterizing the gravitational field are selected from
several available options. The most frequent two choices are the dyad
components of the Weyl spinor in the NP formalism, and the components of the
perturbed metrics. These two descriptors of the state are related to each
other by the second-order differential equations embodying the field equations
and the gauge conditions. Rather than investigating these complicated
relations, we want to compare here the two choices of the observables given by
the dyad components of the Weyl spinor and by the principal null directions.
These alternatives are connected by the quartic equation
\begin{equation}
\Psi_{4}z^{4}+4\Psi_{3}z^{3}+6\Psi_{2}z^{2}+4\Psi_{1}z+\Psi_{0}=0.
\end{equation}
For the unperturbed black hole, the Weyl spinor components have the form
(\ref{Psis}) and the four solutions for the principal null directions pairwise
coincide: $z_{i}^{(0)}=\left(  0,0,-1/w,-1/w\right)  $. We introduce the
normalized coefficients
\begin{equation}
a_{i}=\left(
\begin{array}
[c]{c}%
4\\
i
\end{array}
\right)  \frac{\Psi_{i}}{\Psi_{4}}%
\end{equation}
for $i=0,...,4.$ As in the previous section, we expand all observables of the
perturbed space-time in powers of the perturbation parameter $\varepsilon$,
\begin{equation}
a_{k}=a_{k}^{(0)}+\varepsilon a_{k}^{(1)}+\mathcal{O}\left(  \varepsilon
^{2}\right)  \qquad for\;k=0,...,3.
\end{equation}
The two sets of observable quantities are now related by the symmetric
expressions
%\begin{eqnarray}
%-a_3 &=&\sum_{i=1}^4z_i,\qquad \qquad a_2=\sum_{\begin{array}{c} i=1  \\ k=i+1
%\end{array}} ^4z_iz_k\qquad  \nonumber \\
%-a_1 &=&\sum_{\begin{array}{c} i=1  \\ k=i+1 \\ l=k+1 \end{array}} ^4z_iz_kz_l,\qquad
%a_0=\prod_{i=1}^4z_i \ .
%\end{eqnarray}%
\begin{align}
-a_{3}  & =\sum_{i=1}^{4}z_{i},\qquad\qquad a_{2}=\sum_{i,k=1,\ i<k}^{4}
z_{i}z_{k}\qquad\nonumber\\
-a_{1}  & =\sum_{i,k,l=1,\ i<k<l}^{4} z_{i}z_{k}z_{l},\qquad a_{0}=\prod
_{i=1}^{4}z_{i} \ .
\end{align}
Inserting here the series expansions of the roots
\begin{align}
z_{1}  & =\varepsilon z_{1}^{(1)}+\mathcal{O}\left(  \varepsilon^{2}\right)
\nonumber\\
z_{2}  & =\varepsilon z_{2}^{(1)}+\mathcal{O}\left(  \varepsilon^{2}\right)
\nonumber\\
z_{3}  & =-\frac1w+\varepsilon z_{3}^{(1)}+\mathcal{O}\left(  \varepsilon
^{2}\right) \\
z_{4}  & =-\frac1w+\varepsilon z_{4}^{(1)}+\mathcal{O}\left(  \varepsilon
^{2}\right) \nonumber
\end{align}
we get the linear relations for the first-order observables
\begin{align}
a_{0}^{(1)}  & =0\label{a}\\
a_{1}^{(1)}  & =\frac1{w^{2}}\left(  z_{1}^{(1)}+z_{2}^{(1)}\right)
\label{symms}\\
a_{2}^{(1)}  & =-2wa_{1}^{(1)}-\frac1w\left(  z_{3}^{(1)}+z_{4}^{(1)}\right)
\\
a_{3}^{(1)}  & =-\left(  z_{1}^{(1)}+z_{2}^{(1)}+z_{3}^{(1)}+z_{4}%
^{(1)}\right)  .
\end{align}
Hence we see that the choice of the principal null directions as the set of
observables is inequivalent to choosing the set of Weyl spinor components. By
Eq. (\ref{a}), for space-times with perturbed principal null directions, the
coefficient $a_{0}$ is perturbed only at the second order in the parameter
$\varepsilon.$ The second-order contribution, however, is nonvanishing and has
the form $a_{0}^{(2)}=$ $\frac1{w^{2}}z_{1}^{(1)}z_{2}^{(1)}.$

Here we have an example of a situation when two choices of the sets of
observable quantities yield inequivalent characterizations of the perturbed
space-time. We then need to decide which of these should be used in the
definition of the perturbed space-time. A significant portion of the existing
literature uses the Weyl tensor components for characterization, with no
regard to the behavior of the alternative set of observables such as the
principal null directions. We argue, however, that two space-times having
finite (or of \ lower-order) differences in some of their observable
quantities should not be considered as exhibiting small perturbations. In the
above situation, the two black hole space-times that have first-order
differences of the curvature component $\Psi_{0}$ do not satisfy the linear
relation (\ref{a}), hence they are inequivalent in the sense that they must
have their corresponding principal null directions pointing at finite angles.

The correct interpretation of the above situation can be inferred from the
behavior of rotational perturbations. When one choice of the set of
observables gives rise to space-times which are characterized by finite
differences in certain observables, then this indicates that pairs of
corresponding quantities with infinitesimal differences will only change in a
higher-order approximation with the appropriate choice of the perturbation
parameter. For rotational perturbations, one might carelessly expand the
diagonal components of the metric with respect to some parameter (such as the
square of the angular velocity) such that the perturbed state has first-order
contributions in the diagonal metric. That this is not the right choice of the
parameter, becomes clear when considering the behavior under reflections.

\section{\textbf{Gauge choices}}

The foregoing discussion did not extend to the problem of choosing the tetrad
gauge. In the exact description of a space-time, \ the choice of a null tetrad
vector along a principal null direction leaves only a small discrete group of
gauge symmetries permuting the principal directions. In the present section we
show that the gauge group in the linear approximation is much larger, and in
fact it has an important geometrical content.

The perturbed tetrad is chosen such that the spinor $o^{A}$ is one of the four
principal spinors of the Weyl curvature:
\begin{equation}
\Psi_{0}=0.\label{pnd}%
\end{equation}
For the unperturbed black hole, two independent choices for the spinor $o^{A}$
are possible, corresponding to the two distinct double principal directions.

In perturbation theory, the solutions of (\ref{pnd}) remain, to some extent,
undetermined in the linear approximation. Rather than having a finite number
of exact solutions of Eq. (\ref{pnd}), there exist infinitely many solutions
of the problem. This can be seen from the expansions (\ref{symms}) of the
symmetric expressions. Given the coefficients $a_{i}^{(1)}$, these equations
have an infinite number of solutions for the roots $z_{i}^{(1)}$. The
multitude of the principal null directions in the linear approximations gives
rise to a group of dyad transformations which leaves condition (\ref{pnd}) unchanged.

The quantity $\Psi_{0}$ transforms under the infinitesimal dyad
transformation
\begin{equation}
o^{A}\rightarrow o^{A}+b\iota^{A},\qquad\iota^{A}\rightarrow\iota
^{A},\label{omtr}%
\end{equation}
as follows:
\begin{equation}
\Psi_{0}\rightarrow\Psi_{0}+4b\Psi_{1}.
\end{equation}
Here $b$ is an arbitrary but small complex multiplier function such that
higher powers of $b$ are negligible. Since the curvature quantity $\Psi_{1}$
itself is small, the spinor $o^{A}$ remains a principal spinor of the
curvature under the transformations (\ref{omtr}).

Gauge symmetries are normally considered as mathematical properties of our
description. A gauge transformation is expected to alter such quantities of
the description as the tetrad frame or a field potential while leaving the
physically measurable quantities (the curvature or the field strength)
invariant. From this point of view, the above setup where we choose a tetrad
along a principal null direction is somewhat unusual. Here a gauge
transformation of the type (\ref{omtr}) amounts to selecting a congruence of
principal null curves. The particular congruence thus selected is that with a
tangent spinor $o^{A}.$ Thus a `gauge change' of the form (\ref{omtr}) amounts
to picking a new family of principal null curves. These families of curves may
have different propagation properties. Hence the choice of the `gauge' has
significant consequences for the resulting behavior in the linear approximation.

One of the earliest contributions to the literature of algebraically special
perturbations by Couch and Newman \cite{Couch} takes advantage of an
especially simple form of the metric, and finds the perturbations of a
Schwarzschild black hole. The linear algebraically special perturbations of
the Kerr black hole are considered by Wald \cite{Wald} in the NP approach.
Wald uses a gauge in which $\Psi_{0}=\Psi_{1}=0$. For this particular pick of
the principal null congruence, both the shear and the first curvature vanish.
Wald's aim is to prove that also $\Psi_{4}$ vanishes, in order to establish
that the knowledge of $\Psi_{0}$ alone suffices for a full description of the
perturbed space-time. He fails to achieve this goal for modes with certain
`algebraically special' frequencies. Note that one should proceed with caution
when using the results of this paper. For example, on p. 1457, one finds this
claim: `\textit{i.e.}, we obtain the linearized version of the Goldberg-Sachs
theorem'. Now the Goldberg-Sachs theorem has been shown not to hold in the
linearized theory \cite{Dain}.

Chandrasekhar further investigates the properties of algebraically special
perturbations \cite{Chandrasekhar}. He concludes his work with several,
apparently technical, but at any rate, unanswered questions. These concern the
relation of the Starobinsky operator to complex conjugation, and to the
special forms of the potential barriers occurring. These questions (he writes)
are sheated in enigmas. A detailed investigation of the algebraically special
frequencies is to be found in \cite{Brink}.

Despite the deceptive simplicity achieved by the choice of the principal null
congruence with $\Psi_{1}=0,$ this approach neither offers a complete
description of the space-time nor can it be easily generalized for the
presence of charge.

\section{Second-order perturbations\label{seco}}

In Section\ 6, we have shown that the component $\Psi_{0}$ of the curvature
does not receive a contribution in the linear approximation. In the light of
existing earlier literature of black hole perurbations, it is an important
question what are the equations governing the first nonvanishing contribution
to the field $\Psi_{0}$.

Our goal in this section is to obtain an uncoupled differential equation for
the second-order function $\Psi_{0}$. To this end, we express the derivatives
$\delta\rho$, $\delta\kappa$, $D\tau$, $D\beta$, $D\mu$ and $D\overline
{\alpha}$ from the NP equations (4.2k), (4.2b), (4.2c), (4.2e), (4.2h) and
from the complex conjugate of (4.2d), respectively. Similarly, we express
$D\Psi_{2}$, $\delta\Psi_{2}$, $D\Psi_{1}$ and $\delta\Psi_{1}$ from the NP
electrovacuum Bianchi identities (A3). Where necessary, second derivatives are
untangled by use of the commutator $\left[  D,\delta\right]  $. Acting with
the commutator $\left[  D,\delta\right]  $ on $\Psi_{1}$, the resulting
equation has the form
\begin{align}
& \left[  \delta\bar{\delta}-D\Delta-4\alpha\delta+2\bar{\alpha}\bar{\delta
}+\left(  4\gamma-\mu\right)  D+\left(  \bar{\rho}+4\rho\right)
\Delta\right.  \nonumber\\
& \left.  -8\alpha\bar{\alpha}+4D\gamma-4\delta\alpha-4\bar{\rho}%
\gamma-16\gamma\rho+4\mu\rho+2\Psi_{2}\right]  \Psi_{0}\\
& =-4\left(  \bar{\delta}-6\alpha\right)  \left(  \sigma\Psi_{1}\right)
+4\left(  \Delta-\bar{\gamma}+\bar{\mu}-5\gamma\right)  \left(  \kappa\Psi
_{1}\right)  -10\Psi_{1}^{2}\ .\nonumber
\end{align}
To second order, the terms in the square bracket on the left can be taken to
have their value in the Kerr-Newman metric since they act on the second-order
quantity $\Psi_{0}$. The terms on the right contain either of the second-order
quantities $\sigma\Psi_{1}$, $\kappa\Psi_{1}$ or $\Psi_{1}^{2}$. The operators
acting on these second-order quantities assume their values at the Kerr-Newman
metric:
\begin{align}
\square_{2}\Psi_{0}  & =2\left\{  -\frac{\sqrt{2}}{\zeta}\left[
ia\sin\vartheta\left(  \frac{\partial}{\partial r}-\frac{\partial}{\partial
t}\right)  +\frac{\partial}{\partial\vartheta}-\frac{i}{\sin\vartheta}%
\frac{\partial}{\partial\varphi}+\frac{3\cos\vartheta}{\sin\vartheta}\right]
\sigma\right.  \\
& \left.  +\left[  \left(  \frac{\mathrm{m}(\zeta+\bar{\zeta})-\mathrm{e}^{2}%
}{\zeta\bar{\zeta}}-1\right)  \left(  \frac{\partial}{\partial r}+\frac
{1}{\zeta}\right)  +2\frac{\partial}{\partial t}-6\frac{\mathrm{m}\bar{\zeta
}-\mathrm{e}^{2}}{\zeta^{2}\bar{\zeta}}\right]  \kappa-5\Psi_{1}^{{}}\right\}
\Psi_{1}^{{}}\nonumber
\end{align}
where
\begin{align}
\square_{s}  & ={\mathbf{\Delta}}^{-s}\frac{\partial}{\partial r}%
{\mathbf{\Delta}}^{s+1}\frac{\partial}{\partial r}+\frac{1}{\sin\vartheta
}\frac{\partial}{\partial\vartheta}\sin\vartheta\frac{\partial}{\partial
\vartheta}+s\left(  1-s\frac{\cos^{2}\vartheta}{\sin^{2}\vartheta}\right)
\nonumber\\
& +\left[  2a\left(  \frac{\partial}{\partial t}-\frac{\partial}{\partial
r}\right)  +\frac{1}{\sin^{2}\vartheta}\left(  \frac{\partial}{\partial
\varphi}+2is\cos\vartheta\right)  \right]  \frac{\partial}{\partial\varphi}\\
& +a^{2}\sin^{2}\vartheta\frac{\partial^{2}}{\partial t^{2}}-2\left[
(r^{2}+a^{2})\frac{\partial}{\partial r}+(s+2)r+ia\cos\vartheta\right]
\frac{\partial}{\partial t}.\nonumber
\end{align}
is the wave operator introduced in \cite{bhpart}.

Given the first-order solution of the perturbation problem, we have an
uncoupled linear differential equation for the unknown function $\Psi_{0}$.
The right-hand side is fully known and is to be treated as a source term.

In the second approximation, the issue of the gauge choice arises again. This
has been investigated in \cite{Lousto} and \cite{Bruni}.

\section{Conclusions}

We have argued above that perturbations of a black hole can be described in a
sense which conforms to generally accepted criteria if the lowest-order
contributions to the curvature component $\Psi_{0}$ are of second order in the
perturbation parameter. First-order contributions to $\Psi_{0}$ would result
in finite changes in the principal null directions, which are directly
accessible to observations. Before these disturbing results can be considered
as fully consolidated, there is a clear need to clarify a number of related
issues. We conclude this work with raising just one such example.

The curvature component $\Psi_{4}$ remains invariant under transformations
(\ref{omtr}). Under infinitesimal dyad transformations of the form
\begin{equation}
\iota^{A}\rightarrow\iota^{A}+co^{A},\qquad o^{A}\rightarrow o^{A}%
,\label{iotr}%
\end{equation}
with $c$ an arbitrary small complex function, $\Psi_{4}$ transforms
\begin{equation}
\Psi_{4}\rightarrow\Psi_{4}+4c\Psi_{3}.
\end{equation}
In the Kinnersley tetrad, the spinor $\iota^{A}$ is chosen, as is $o^{A}$, to
be a principal spinor for the Kerr metric \cite{Kinnersley}, and both
$\Psi_{3} $ and $\Psi_{4}$ vanish. Repeating the argument with the symmetrical
expressions (\ref{symms}) for $\Psi_{4}^{(1)}$ and projective coordinate
$z^{\prime}=-1/z$, at first sight it would seem to be possible to choose the
null tetrad for the perturbed space-time such that the only non-vanishing
tetrad component of the Weyl spinor is again $\Psi_{2}$. This would imply that
the perturbed space-time is again the Kerr black hole with trivial parameter changes.

To find ways out from controversies like the one just described, either one
has to give up insisting on the observability of principal null directions, or
else to carry out a careful investigation of perturbations in a neighborhood
of the coordinate singularity on the two-sphere. We do not see any ground,
however, on which the observability of principal null directions should be
given up.

\section*{Acknowledgments}

Correspondence on various aspects of this work with colleagues is herewith
acknowledged. We thank Hisaaki Shinkai for literature on observational
techniques, Lior Burko for sharing his views on the algebraic type, Amos Ori
for advice on the scaling of the Weyl tensor components under perturbations,
and Sergio Dain on the Goldberg-Sachs theorem. This work has been supported by
the OTKA grant T031724.


\begin{thebibliography}{99}                                                                                               %
\bibitem {Kramer}M.M.D. Kramer, M.A.H. MacCallum, H. Stephani and E. Herlt:
\textit{Exact Solutions of Einstein's Field Equations} (Cambridge University
Press, Cambridge, 1980).

\bibitem {NP}E.T. Newman and R. Penrose, J. Math. Phys. \textbf{3}, 566
(1962), to be referred to as NP.

\bibitem {Baker}J. Baker and M. Campanelli, Phys. Rev. \textbf{D62}, 127501 (2000).

\bibitem {Kinnersley}W. Kinnersley, J. Math. Phys. \textbf{10}, 1195 (1969).

\bibitem {Couch}W.E. Couch and E.T. Newman, J. Math. Phys. \textbf{14}, 285 (1973).

\bibitem {Wald}R.M. Wald, J. Math. Phys. \textbf{14}, 1453 (1973).

\bibitem {Chandrasekhar}S. Chandrasekhar, Proc. Roy. Soc. \textbf{A392}, 1 (1984).

\bibitem {Brink}A.M. van den Brink, Phys. Rev. \textbf{D62}, 064009 (2000),
gr-qc/0001032 preprint (2000).

\bibitem {Stewart}J.M. Stewart and M. Walker, Proc. R. Soc. London
\textbf{A341}, 49 (1974).

\bibitem {Chandra2}S. Chandrasekhar: \textit{The Mathematical Theory of Black
Holes} (Clarendon Press, 1983, p. 430).

\bibitem {Karlhede}A. Karlhede, Gen. Rel. Gravitation \textbf{12}, 693 (1980).

\bibitem {Michael}M. Bradley, in \textit{Relativity Today}, Eds. C.A.
Hoenselaers and Z. Perj\'{e}s (Akad\'{e}miai Kiad\'{o}, 2002).

\bibitem {Hartle}J. Hartle, Astrophys. J. \textbf{150}, 1005 (1967).

\bibitem {PR}R. Penrose and W. Rindler: \textit{Spinors and Space-Time}, Sec.
8 (Cambridge University Press, Cambridge, 1986).

\bibitem {Gunnarsen}L. Gunnarsen, H. Shinkai and K. Maeda, Class. Quantum
Grav. \textbf{12}, 133 (1995), gr-qc/9406003 preprint (1994).

\bibitem {Kristian}J. Kristian and R.K. Sachs, Astrophys. J. \textbf{143}, 379 (1965).

\bibitem {Chrobok}T. Chrobok and V. Perlick, Class. Quantum Grav. \textbf{18},
3059 (2001).

\bibitem {MTW}C. Misner, K.S. Thorne and J.A. Wheeler: \textit{Gravitation}
(Freeman, 1973).

\bibitem {Dain}S. Dain and O.M. Moreschi, J. Math. Phys. \textbf{41}, 6296
(2000), gr-qc/0203057 preprint (2002).

\bibitem {bhpart}Z. Perj\'{e}s and M. Vas\'{u}th, Astrophys. J. \textbf{582},
342 (2003).

\bibitem {Lousto}M. Campanelli and C.O. Lousto, Phys. Rev. \textbf{D59, }
124022 (1999).

\bibitem {Bruni}M. Bruni, S. Mattarrese, S. Mollerach and S. Sonego, Class.
Quantum Grav. \textbf{14}, 2585 (1997), gr-qc/9609040 (1996).
\end{thebibliography}
\end{document}